\documentclass[prl,aps,reprint,noeprint,superscriptaddress,amsmath, amssymb]{revtex4-1}
\usepackage{graphicx}
\usepackage{subfigure}
\usepackage{lineno}
\usepackage{epsfig}
\usepackage{multirow}
\usepackage{overpic}
\usepackage{color}
\usepackage{bm}
\usepackage{xspace} 
\usepackage{dcolumn}% Align table columns on decimal point
\usepackage{booktabs,tabularx}
\usepackage{array}
\usepackage{textcomp}
\usepackage[colorlinks,linkcolor=blue,urlcolor=blue,citecolor=blue]{hyperref}
\usepackage{tikz}

\usepackage{cleveref}%multiple figure reference

\newcommand{\EE}{e^+e^-}

\newcommand{\SSm}{\Sigma^-\bar\Sigma^+}
\newcommand{\SSp}{\Sigma^+\bar\Sigma^-}
\newcommand{\SSz}{\Sigma^0\bar\Sigma^0}

\begin{document}

\normalsize
\parskip=5pt plus 1pt minus 1pt

%%%%%%%%%%%%%%%%%%%%%%%%%%%%%%%%%%%%%%%%%%%%%%%%%%%%%%%%%%%%%%%%%

\title{
\boldmath Measurement of Born cross sections for $\EE\to\SSm$ at $\sqrt{s}=3.51-4.95$ GeV and observation of $\psi(3770)\to\Sigma^-\bar\Sigma^+$ 
}
\author{BESIII Collaboration}
\thanks{Full author list given at the end of the Letter.}
%\author{\small\input{authorlist_2025-09-11}}
%%%%%%%%%%%%%%%%%%%%%%%%%%%%%%%%%%%%%%%%%%%%%%%%%%%%%%%%%%%%%%%%%
\begin{abstract}
Using $e^+e^-$ collision data corresponding to an integrated luminosity of 44 fb$^{-1}$ collected with the BESIII detector at the BEPCII Collider, we report the first measurement of Born cross sections and effective form factors for $\EE\to\SSm$ at center-of-mass energies between 3.51 and 4.95 GeV. With a fit to the $\sqrt{s}$-dependent dressed cross sections, the decay $\psi(3770)\to\SSm$ is observed for the first time with a significance of 5.5$\sigma$, including systematic uncertainties. 
The corresponding branching fraction is determined to have two solutions, $(2.9 \pm 0.3)\times 10^{-4}$ and $(1.1 \pm 0.1)\times 10^{-4}$, where the quoted uncertainties also include systematic uncertainties.
Upper limits at the 90\% confidence level on the product of the branching fraction and the electronic partial width are provided for other possible charmonium(like) states.
The ratios of Born cross sections for the $\Sigma$ isospin-triplet states are determined and can be used to test the vector-meson-dominance model.
\end{abstract}

\newcommand{\BESIIIorcid}[1]{\href{https://orcid.org/#1}{\hspace*{0.1em}\raisebox{-0.45ex}{\includegraphics[width=1em]{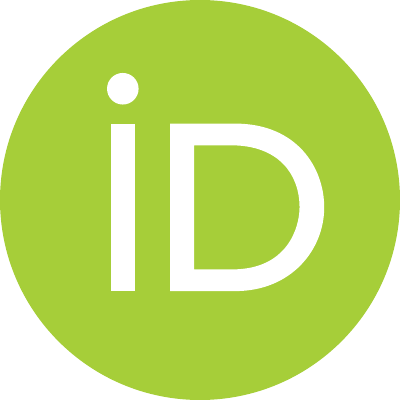}}}}
\maketitle
%\linenumbers

%%%%%%%%%%%%%%%%%%%%%%%%%%%%%%%%%%%%%%%%%%%%%%%%%%%%%%%%%%%%%%%%%
%Introduction
%%%%%%%%%%%%%%%%%%%%%%%%%%%%%%%%%%%%%%%%%%%%%%%%%%%%%%%%%%%%%%%%%%%%%%%%%%%%%%%
Since the discovery of the first charmonium state, $J/\psi$, in 1974~\cite{E598:1974sol,SLAC-SP-017:1974ind}, numerous charmonium states have been observed. Above the open-charm threshold, the potential-quark model~\cite{Barnes:2005pb} predicts six states with masses between 3.7 and 4.7 GeV, corresponding to 1D, 3S, 2D, 4S, 3D, and 5S configurations~\cite{Brambilla:2010cs}. Experimentally, however, an excess of vector states has been detected in this region. The 3S, 2D, and 4S states have been identified as $\psi(4040)$, $\psi(4160)$, and $\psi(4415)$ through inclusive hadronic cross sections~\cite{BES:2001ckj}. In addition, four unconventional states—$Y(4230)$, $Y(4360)$, $Y(4634)$, and $Y(4660)$—have been observed in hidden-charm final states by BaBar, Belle~\cite{BaBar:2005hhc,BaBar:2006ait,Belle:2007umv,Belle:2007dxy,Belle:2008xmh,BaBar:2012vyb,BaBar:2012hpr,Belle:2013yex,Belle:2014wyt}, CLEO~\cite{CLEO:2006ike}, and BESIII~\cite{BESIII:2014rja,BESIII:2023cmv}. These $Y$ states are unlikely to be pure $c\bar{c}$ resonances; their overpopulation and the mismatch between predicted and measured properties make them prime candidates for exotic hadrons. Proposed explanations include hybrid, multiquark, or molecular configurations~\cite{Brambilla:2010cs,Close:2005iz,Briceno:2015rlt,Chen:2016qju,Wang:2019mhs,Qian:2021neg,Dai:2023vsw,Wang:2023zxj,Wang:2025dur,Bai:2026atm}. A series of experimental measurements have been conducted to investigate these hypotheses~\cite{BESIII:2015aym,BESIII:2017nty,Belle:2018jqa,BESIII:2023tll}, but no definitive conclusions have been reached yet.
These states provide an ideal laboratory for exploring nonperturbative QCD dynamics, thereby offering important insights into fundamental aspects of the strong interaction, such as quark confinement and the role of gluonic fields, and advancing our understanding of matter at the subatomic scale.

Hyperon-pair production in $e^+e^-$ annihilation or charmonium(like) decays offers a clean final-state topology and provides direct access to the underlying interaction mechanisms, assumed to proceed via one-photon or three-gluon exchange. Consequently, two-body hyperonic decays of charmonium(like) states above the open-charm threshold are a powerful tool for elucidating their nature~\cite{Close:2005iz,Wang:2019mhs,Qian:2021neg}. In particular, hybrid models predict charmless decays of charmoniumlike states~\cite{Close:2005iz}, while theories that describe these states as mixtures of conventional charmonium states also require experimental verification via their possible decays into light-hadron final states~\cite{Wang:2019mhs,Qian:2021neg}.
Although extensive searches have been conducted by BESIII and Belle, only $\psi(3770)\to\Lambda\bar\Lambda$~\cite{BESIII:2021ccp} and $\psi(3770)\to\Xi^-\bar\Xi^+$~\cite{BESIII:2023rse} have been observed so far.

Furthermore, the measured Born cross sections can be used to determine the hyperon effective form factors (EFFs) $G_{\rm eff}(s^2)$, which are functions of the $\gamma^*$-baryon-antibaryon vertex described by two independent electromagnetic form factors (EMFFs), $G_E$ and $G_M$~\cite{Lorenz:2015pba,Mangoni:2021qmd}. In the timelike region, the EMFFs can be related to the time evolution of the internal charge and magnetic distributions of the hyperon and then offer insight into the internal structure of hyperon. The vector-meson-dominance (VMD) model successfully describes hyperon EMFFs~\cite{Iachello:1972nu,Iachello:2004aq,Bijker:2004yu}, as demonstrated for $\Lambda$~\cite{Yang:2019mzq,Li:2021lvs}, $\Sigma$~\cite{Li:2020lsb,Dai:2021yqr,Yan:2023yff}, and $\Xi$~\cite{Dai:2021yqr,Yan:2023yff} near production threshold. 
As emphasised in Ref.~\cite{Yan:2023yff}, the study of $\EE\to\SSm$ Born cross sections and EFFs will complete the measurement of the $\Sigma$ isospin triplet~\cite{BESIII:2024umc,BESIII:2024gql}, enabling a comprehensive test of the VMD model and thereby providing a stringent validation of its applicability.
%As emphasised in Ref.~\cite{Yan:2023yff}, ratios of Born cross sections and EFFs between $\EE\to\SSm$ and its isospin partners $\EE\to\SSp$ and $\EE\to\SSz$ provide stringent tests of the VMD model.

In this Letter, we present the first measurement of the Born cross sections and the EFFs for the $\EE\to\SSm$ reaction at 52 center-of-mass (c.m.) energies ($\sqrt{s}$) between 3.51 and 4.95 GeV, using a total integrated luminosity of 44 $\rm fb^{-1}$ collected with the BESIII detector at the BEPCII collider.
Details of BEPCII and BESIII are given in Refs.~\cite{Ablikim2010, BESIII:2020nme,Yu:2016cof, Huang:2022wuo, Li:2017eToF, Guo:2017eToF, Cao:2020ibk}. The potential resonances are searched by fitting the $\sqrt{s}$-dependent dressed cross sections of the $\EE\to\SSm$ reaction, and the $\psi(3770)\to\Sigma^-\bar\Sigma^+$ decay is observed, leading to the first observation of $\psi(3770)\to\SSm$.

Signal candidates of $\EE\to\SSm$ are selected via $\Sigma^- \to n \pi^-$ and $\bar\Sigma^+ \to \bar{n} \pi^+$, involving one neutron and one antineutron in the final state. 
The neutron is not required in the selection to avoid photon contamination due to their similar features in the electromagnetic calorimeter (EMC), and
only the antineutron is required to be detected due to its higher energy deposition as antimatter in the EMC compared to the neutron.
Therefore, a partial reconstruction technique with a missing neutron is employed in this work.
Monte Carlo (MC) simulated data samples are produced with a {\sc geant4}-based~\cite{G42002iii} software package, which includes the geometric description of the BESIII detector and the detector response. They are used to determine detection efficiencies and estimate backgrounds. The simulation models the beam energy spread and initial state radiation (ISR) in the $e^+e^-$ annihilations with the generator {\sc kkmc}~\cite{Jadach2001}. 
The exclusive MC samples for the signal process are generated using a phase space (PHSP) model implemented in the EvtGen event generator~\cite{Lange:2001uf,Ping:2008zz}, with 100,000 events simulated at each energy point.
To examine potential background channels, an inclusive MC sample of $e^+e^-\to$ hadrons at $\sqrt{s} =3.773$~GeV is analyzed, which includes $D\bar{D}$ decays (with quantum coherence for neutral $D$ modes), non-$D\bar{D}$ decays of $\psi(3770)$, ISR production of $J/\psi$ and $\psi(3686)$, and continuum processes implemented in {\sc kkmc}~\cite{Jadach2001}. 

% \section{Event selection}
Charged tracks must be reconstructed in the main drift chamber (MDC) within the angular region $|\cos\theta|<0.93$, where $\theta$ is the polar angle with respect to the symmetry axis of the MDC (the $z$ axis) in the laboratory frame.
The distance of closest approach to the interaction point must be less than 30 cm along the $z$ axis, and less than 10~cm in the transverse plane. The number of positive and negative charged tracks is required to be one. 
Particle identification for charged tracks combines the measurements of the specific ionization energy loss in the MDC (${\rm d}E/{\rm d}x$) and the time of flight to form likelihoods $\mathcal{L}(h)$ $(h=p,K,\pi)$ for each hadron ($h$) hypothesis. 
Tracks are identified as pions when the pion hypothesis has the greatest likelihood $(\mathcal{L}(\pi)>\mathcal{L}(K)~\&\&~\mathcal{L}(\pi)>\mathcal{L}(p))$. 
The events with only one $\pi^+$ and one $\pi^-$ are kept for further analysis. 
To remove the cosmic ray background, the difference in the times of flight between the $\pi^+$ and $\pi^-$ within the MDC must be greater than $-5$ ns.

Antineutrons are reconstructed and identified using EMC showers. Based on the EMC energy resolution, the initial requirement on the deposited energy of each shower is set to be greater than 25~MeV in the barrel region ($|\cos\theta| < 0.80$) and 50~MeV in the end-cap region ($0.86<|\cos\theta|<0.92$).  
To suppress electronic noise and showers unrelated to the event, the difference between the EMC time of the neutral candidate and the event start time is required to be within [0, 700]~ns. 
The shower with the highest deposited energy is identified as an antineutron. 
Based on MC studies, the requirement for the antineutron deposited energy is set to be greater than 0.62~GeV at $\sqrt{s} = 3.773$~GeV and varies for different energy points, to exclude misidentified neutron events.  The remaining misidentified neutron events contribute a fraction of below 0.1\%, do not form a peaking background, and thus can be neglected.
The second moment of the shower, $S_{\bar{n}}=\sum_i E_i r_i^2/\sum_i E_i$, is required to exceed 18 cm$^2$, where $E_i$ is the energy deposited in crystal $i$, and $r_i$ is its distance from the shower center. The total number of EMC hits associated with the shower must exceed 16. 
To reject radiative backgrounds and showers associated with charged tracks, the angle between the antineutron candidate shower and the extrapolated position of the nearest charged track at the EMC entrance must exceed 15 degrees. 

To suppress the background and enhance the mass resolution, a one-constraint (1C) kinematic fit is employed, and the goodness of fit, $\chi^2_{\rm 1C}$, has to be less than 60.
Energy–momentum conservation provides four constraints, while the $\pi^+\bar{n}$ combination is constrained to the known $\bar\Sigma^+$ mass~\cite{ParticleDataGroup:2024cfk}. 
Because the neutron four-momentum is unavailable, only its mass is constrained; for the antineutron, both mass and flight direction are constrained.
To further suppress the backgrounds, the signal region on the recoiling mass $M_{\bar{n}\pi^+}^{\rm Recoil}$ is required to be within the interval $[1.10, 1.35]$~GeV/$c^2$.
% \section{Background study and signal yields}
After applying the above event selection criteria and analyzing the inclusive MC samples at $\sqrt{s} = 3.773$ GeV, the surviving backgrounds are found to distribute smoothly in the signal region.

\begin{figure}[htbp]
    %\begin{center}
    \hspace{0.8cm}
    \includegraphics[width=0.45\textwidth]{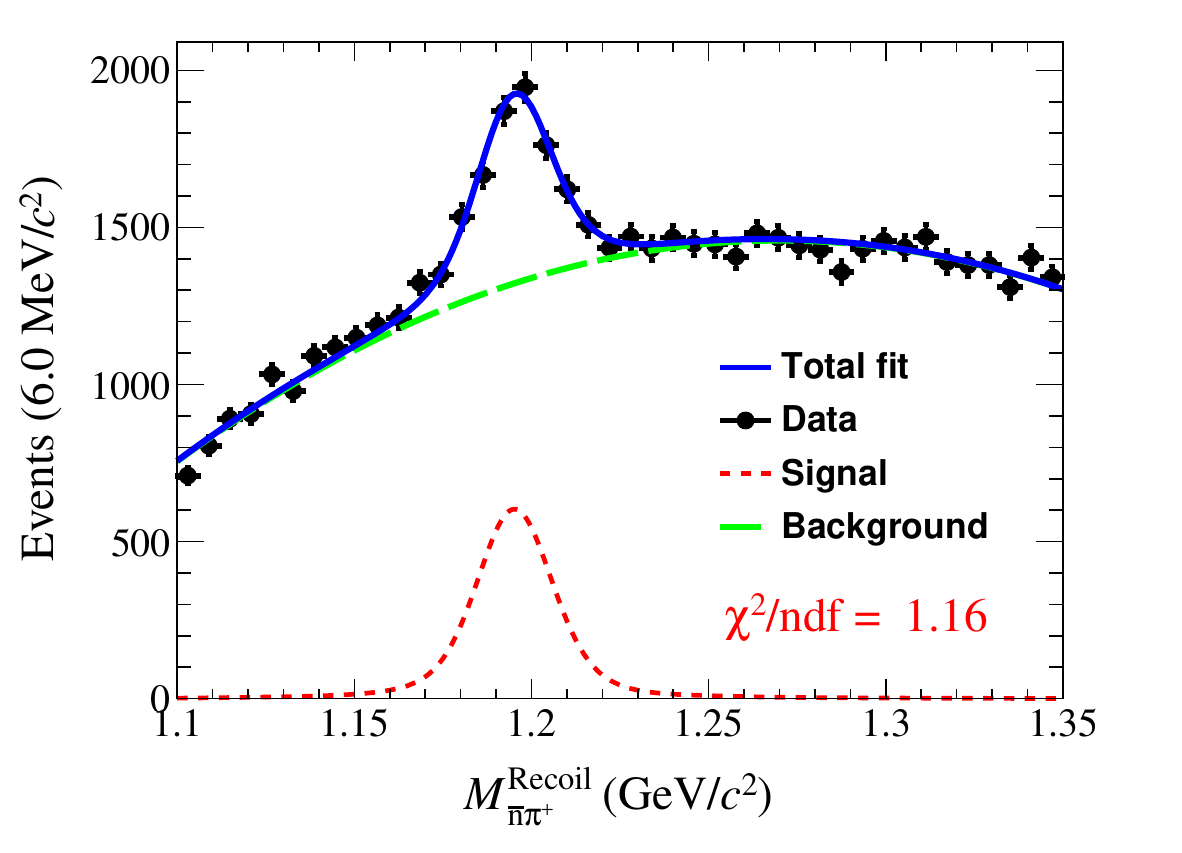}
    \hspace{-0.8cm}
    %\end{center}
    \vspace{-0.5cm}
    \caption{Fit to the $M_{\bar{n}\pi^+}^{\rm Recoil}$ distribution at $\sqrt{s}=3.773$ GeV. The dots with error bars are data, the blue solid line represents the fit result, the red dashed line is for signal shape, and the green dashed line stands for the background.}
    \label{3773massfit}
\end{figure}

To extract the signal yield for the process $\EE\to\SSm$,
an extended maximum likelihood fit to the $M_{\bar{n}\pi^+}^{\rm Recoil}$ distribution in the range from 1.1 to 1.35~GeV/$c^2$ is performed for each point of the energy scan. In the fit, the signal shape of the $M_{\bar{n}\pi^+}^{\rm Recoil}$ spectrum is represented by the simulated MC shape convolved with a Gaussian function to account for the mass resolution difference between the data and the MC simulation at $\sqrt{s} = 3.773$ GeV. For other energy points, only the simulated MC shape is applied in the fit due to the limited statistics. The impact of using different signal shapes is evaluated as a source of systematic uncertainty.
Backgrounds are described by a second-order Chebyshev polynomial. Figure~\ref{3773massfit} shows the fit result for the data at $\sqrt{s} = 3.773$~GeV.

% \section{BCS and EFF}
The Born cross section for the $\EE\to\SSm$ process at a given c.m.\ energy is calculated as
\begin{equation}
\sigma^{B} =\frac{N_{\rm obs}}{{\cal{L}}\cdot(1 + \delta)\cdot\frac{1}{|1 - \Pi|^{2}}\cdot\epsilon\cdot {\cal B}^2_{\Sigma^-\to n\pi^-}},
\end{equation}
where $N_{\rm obs}$ is the signal yield, ${\cal{L}}$ is the integrated luminosity, $(1 + \delta)$ is the ISR correction factor, $\frac{1}{|1-\Pi|^2}$ is the vacuum polarization (VP) correction factor, $\epsilon$ is the detection efficiency, and the ${\cal B}_{\Sigma^-\to n\pi^-}$ is the corresponding branching fraction taken from the Particle Data Group (PDG)~\cite{ParticleDataGroup:2024cfk}. 
The ISR correction factor is obtained by using the QED calculation as described in Ref.~\cite{Kuraev:1985hb}. The VP correction factor is calculated according to Ref.~\cite{Jegerlehner:2011ti}. 
Note that considering the difference in detection efficiency between MC and data, a correction for MC efficiency is employed based on the final states $\pi^-$, $\pi^+$, and $\bar{n}$, using the same correction algorithm and factors in Ref.~\cite{BESIII:2022upt}, and the efficiency difference caused by incorporating the $\bar{n}$ EMC information in the kinematic fit is also considered.
The efficiencies and ISR correction factors are determined through an iterative procedure to accurately measure the Born cross section as proposed in Ref.~\cite{Sun:2020ehv}.

\begin{figure}[!hbpt]
	%\begin{center}
        \hspace{0.6cm}
        \includegraphics[width=0.45\textwidth]{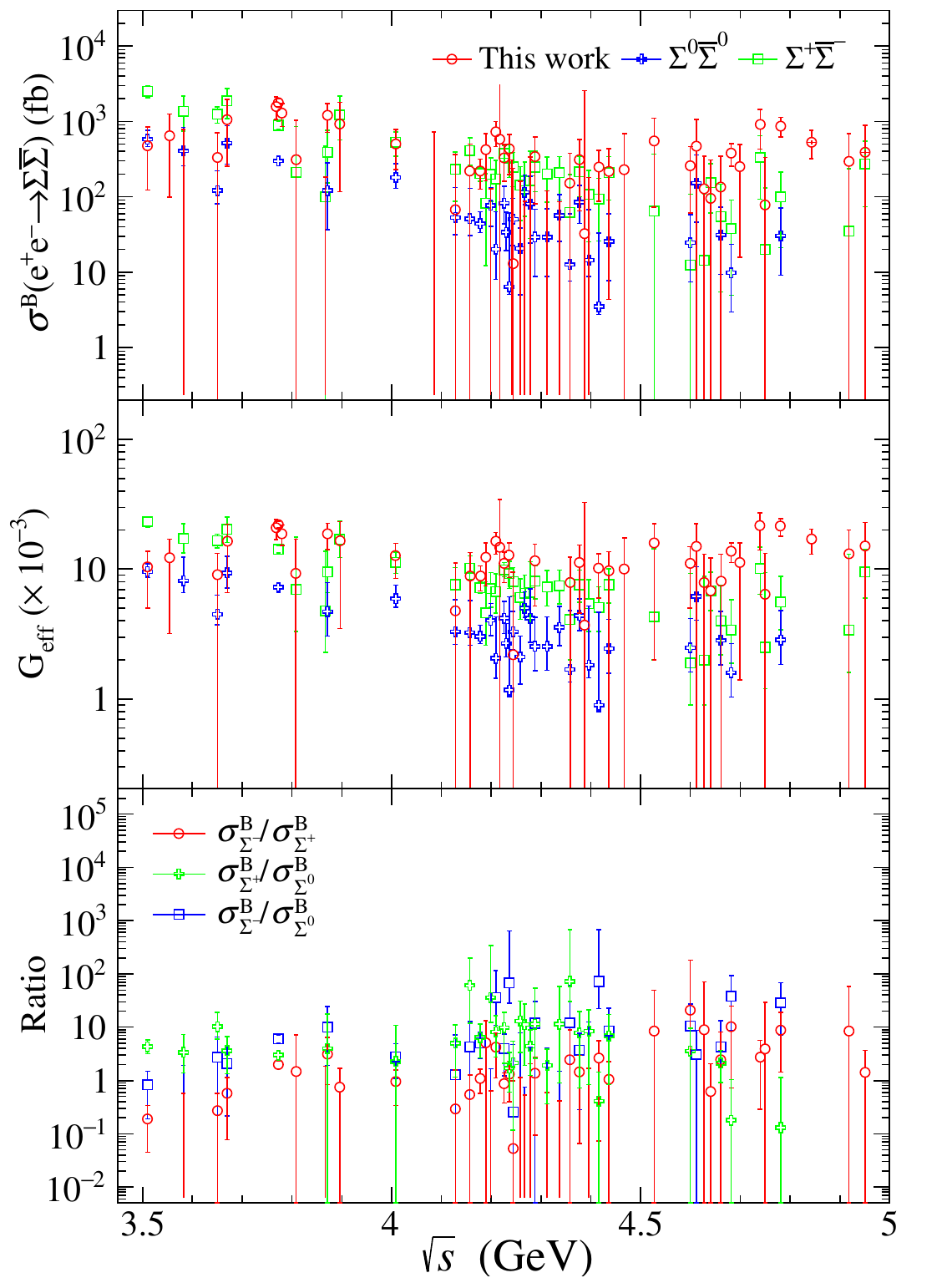}
        \hspace{-0.6cm}
	%\end{center}

	\caption{The Born cross sections (top) and the EFFs (middle) as a function of the c.m.\ energy for the $\EE\to\SSm$ channel, in comparison to the previous measurements for $\EE\to\SSz$ \cite{BESIII:2024gql} and $\EE\to\SSp$ \cite{BESIII:2024umc}. The bottom plot shows the ratios of Born cross sections between the $\EE\to\SSm$ (this work), $\EE\to\SSz$, and $\EE\to\SSp$ processes.}
	\label{com_bcs_eff}
\end{figure}

The $\Sigma^-$ EFF defined in Ref.~\cite{Wang:2022zyc},
\begin{equation}
        |G_{\rm eff}(s)| = \sqrt{\frac{3s\sigma^B}{4\pi\alpha^2C\beta(\frac{1}{2\tau}+1)}},
\end{equation}
is determined, where  $\alpha = \frac{1}{137}$ is the fine structure constant, the variable $\beta = \sqrt{1-\frac{1}{\tau}}$ is the velocity, $\tau = \frac{s}{4m_{\Sigma^-}^2}$ with the invariant mass of $\Sigma^-$ particle, $m_{\Sigma^-}$ \cite{ParticleDataGroup:2024cfk}, and the Coulomb factor $C$~\cite{Baldini:2007qg, Arbuzov:2011ff} parameterizing the electromagnetic interaction between the outgoing hyperon and antihyperon.
Figure~\ref{com_bcs_eff} shows the line shapes of the measured Born cross sections and EFFs for $\EE\to\SSm$, $\EE\to\SSp$~\cite{BESIII:2024umc}, and $\EE\to\SSz$~\cite{BESIII:2024gql}. The cross section ratios for these three processes are also provided to validate the predictions based on the VMD model~\cite{Iachello:1972nu,Iachello:2004aq,Bijker:2004yu, Yang:2019mzq,Li:2021lvs}. 
Except for the region around $\sqrt{s} = 3.773$ GeV, where strong interaction contributions may be present, the results at the other energy points are consistent with the VMD predictions within 1 to 2 standard deviations.
The numerical results for each c.m.\ energy are summarized in the Supplemental Material~\cite{Supp:Mat}.
%\section{systematic uncertianty}

The systematic uncertainty on the measurement of the Born cross section mainly arises from the integrated luminosity, branching fractions of the intermediate states, efficiency correction for pions and $\bar{n}$, signal and background parametrizations, physical model, kinematic fit, and input line shape for the cross section. 
The uncertainty due to the integrated luminosity is evaluated in ~\cite{BESIII:2015qfd, BESIII:2022dxl, BESIII:2022ulv}. 
The uncertainty due to the branching fractions of the intermediate states $\Sigma^-\to n\pi^-$ and $\bar\Sigma^+\to\bar{n}\pi^+$ is taken from the PDG~\cite{ParticleDataGroup:2024cfk} and found to be negligible.
%The uncertainties associated with the correction of the selection efficiency of pions and $\bar{n}$ are evaluated with the same method in Ref.~\cite{BESIII:2022upt}.
The uncertainties associated with the correction of the selection efficiency of pions and $\bar{n}$ are evaluated by varying the correction factors within $1\sigma$, as detailed in Ref.~\cite{BESIII:2022upt}.
The uncertainty of the signal shape for the fit of $M^{\rm Recoil}_{\bar{n}\pi^+}$ is evaluated by combining all energy points and varying the Gaussian parameters within a range of 1$\sigma$, and the yield difference is taken as the systematic uncertainty. Moreover, the uncertainty due to the background shape is estimated through an alternative fit with a third-order polynomial function. The resulting difference between the nominal and changed ones is taken as the systematic uncertainty.
The uncertainty for the dependence on the physical model is assessed by comparing efficiencies obtained with the PHSP generator and with the angular distribution measured in the high-statistics control sample of $e^+e^-\to\Sigma^-\bar\Sigma^+$ at $\sqrt{s}=3.773$ GeV. %Measurement of the angular distribution parameter is performed with the same method introduced in Ref.~\cite{BESIII:2022upt}. 
The measurement of the angular distribution parameter $\alpha_{\Sigma^-}$ is performed by fitting the $\cos^2\theta_{\Sigma^-}$ distribution with the function $1+\alpha_{\Sigma^-}\cos^2\theta_{\Sigma^-}$, as detailed in Ref.~\cite{BESIII:2022upt}, where $\theta_{\Sigma^-}$ is the polar angle of the $\Sigma^-$. The fitted value is $\alpha_{\Sigma^-} = 0.97 \pm 0.42$, with the uncertainty being statistical only.
Here, we vary the measured value of the angular distribution parameter by $\pm 1\sigma$ to evaluate the systematic uncertainty for this source. 
The uncertainty due to the kinematic fit includes two parts: the correction of the track helix parameters and the EMC information of the $\bar{n}$ incorporated in the fit. The first part is evaluated with or without the correction, and the second part is estimated by the same method of efficiency correction for pion and $\bar{n}$ selection mentioned above. The total systematic uncertainty due to the kinematic fit is calculated by adding both contributions in quadrature.
The input cross section line shape is determined by the least-squares fit described below, and its systematic uncertainty consists of two components: the input line shape parametrization and the model of the input line shape. For the first component, 200 sets of parameters are sampled according to the parameter covariance matrix, and the ratio of the fitted Gaussian width to its mean is taken as the corresponding systematic uncertainty. For the second component, the systematic uncertainty is evaluated using the $\psi(3770)$ fit with the highest significance described below, by varying the fixed resonance parameters within 1$\sigma$, as described in Ref.~\cite{BESIII:2024gql}. The total value is summed by adding both ones in quadrature. 
The systematic uncertainties for the measured cross section are summarized in the Supplemental Material~\cite{Supp:Mat}, and the total systematic uncertainty is calculated by adding the individual contributions in quadrature.

%\section{fitting}
Resonant contributions to the $\EE\to\SSm$ reaction are investigated by fitting the line shape of dressed cross sections, determined by $\sigma^{\rm dressed}=\sigma^{B}/|1-\Pi|^2$, thereby incorporating the VP effect. The fit is performed using a least-squares method, $\chi^2=\Delta X^TV^{-1}\Delta X$.
This is accomplished by taking into account the covariance matrix $V$ and the vector of residuals $\Delta X$ between the measured and fitted cross sections. 
The covariance matrix incorporates both the correlated and uncorrelated uncertainties among different c.m.\ energies. The systematic uncertainties associated with the luminosity, branching fraction, detection efficiency, and kinematic fit are assumed to be fully correlated among the c.m.\ energies, while the other systematic uncertainties are assumed to be uncorrelated.

\begin{figure}[!hbpt]
	\begin{center}
        \includegraphics[width=0.4\textwidth]{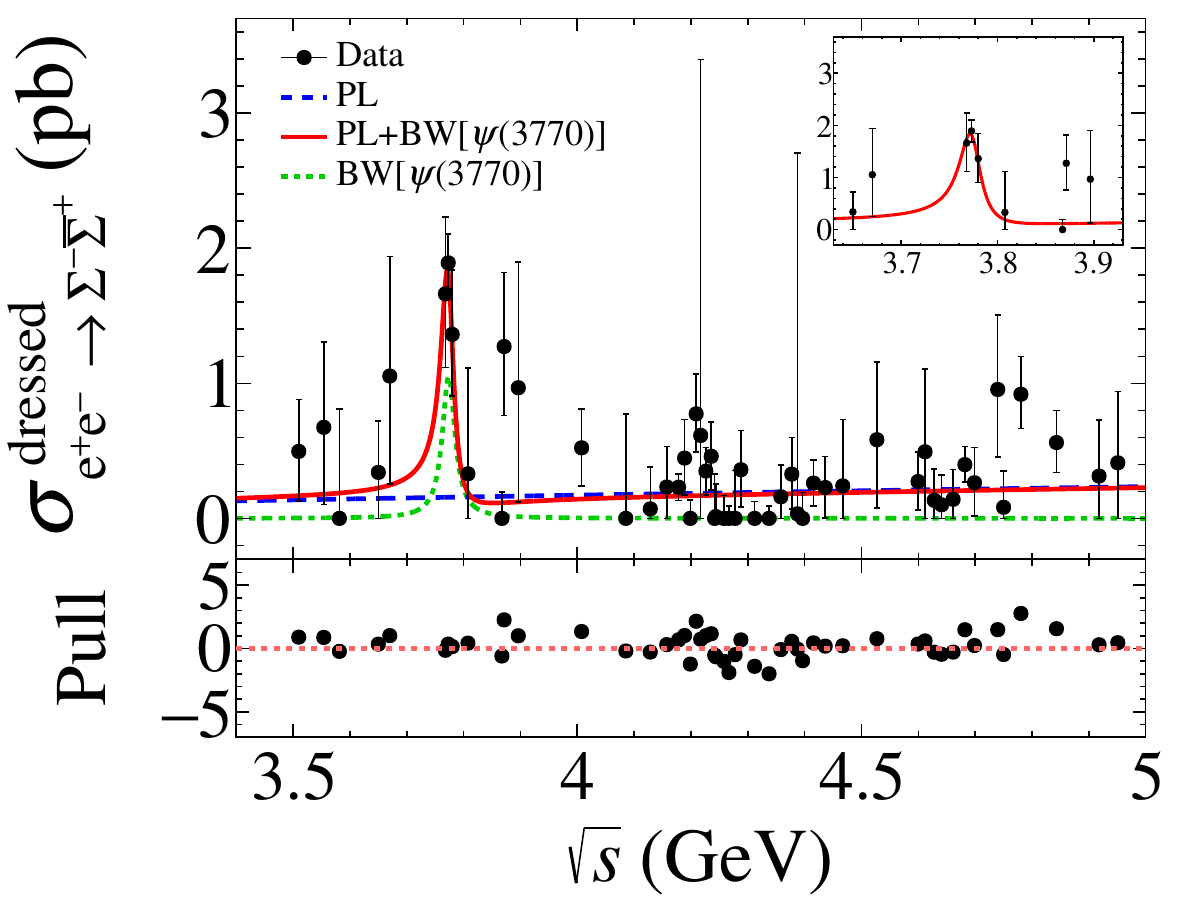}\\\makebox[0.485\textwidth][r]{%
        \includegraphics[width=0.45\textwidth]{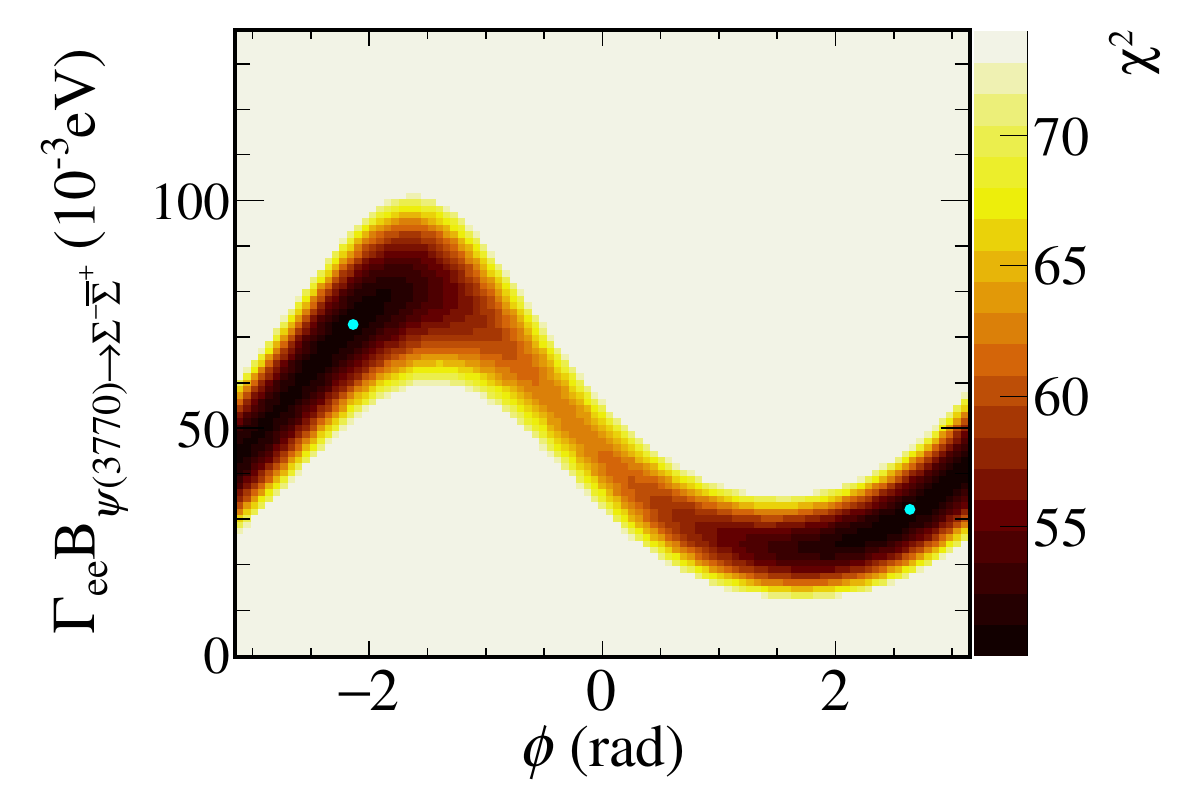}}
	\end{center}
        \vspace{-0.5cm}
	\caption{Fit to the dressed cross sections of $\EE\to\SSm$ (top). The dots with error bars are the measured dressed cross sections including the statistical and systematic uncertainties. The blue dashed line corresponds to the fit curve with only the PL function, while the green dashed line represents the signal of $\psi(3770)$ resonance. The solid red line denotes the fit after incorporating both contributions. The inset plot shows a zoomed-in view around 3.773 GeV to better distinguish the nearby points. The contour of $\Gamma_{ee}\mathcal{B}$ and $\phi$ on the distribution of $\chi^2$ values (bottom). The cyan points represent the two solutions from the best fit.}
	\label{Fig:XiXi::CS::Line-shape-3773}
\end{figure}

Assuming the line shape of cross sections of $\EE\to\SSm$ is formed by a continuum contribution plus a resonance, a fit to the dressed cross sections is applied with the coherent sum of a power-law (PL) function plus a Breit-Wigner (BW) function,
\begin{equation}\label{BCS_1}
	\sigma^{\rm dressed}(\sqrt{s})= \left|c_{0}\frac{\sqrt{P(\sqrt{s})}}{\sqrt{s}^{n}} + e^{i\phi}{\rm BW}(\sqrt{s})\sqrt{\frac{P(\sqrt{s})}{P(M)}}\right|^{2},
\end{equation}
where ${\rm BW}(\sqrt{s}) =\frac{\sqrt{12\pi\Gamma_{ee}{\cal{B}}\Gamma}}{s-M^{2}+iM\Gamma}$, $\phi$ represents the relative phase between the BW function
and the PL function. The $c_0$ and $n$ are free fit parameters, and $\sqrt{P(\sqrt{s})}$ is the two-body PHSP factor. The mass $M$ and total width $\Gamma$ of the assumed resonance are fixed to individual PDG values~\cite{ParticleDataGroup:2024cfk}, and $\Gamma_{ee}{\cal{B}}$ is the product of the electronic partial width and the branching fraction for the assumed resonance decaying into the $\SSm$ final state. 
The fit results with different resonance hypotheses are summarized in Table~\ref{tab:multisolution}. 
Figure~\ref{Fig:XiXi::CS::Line-shape-3773} shows the fit to the dressed cross section using the PL function, the $\psi(3770)$ BW component, and their combination, where a clear $\psi(3770)$ signal is observed. The scan for multiple solutions is also shown.

Multiple solutions arise from the quadratic form of Eq.~(\ref{BCS_1}).  They are obtained by scanning the parameter space and are summarized in the Supplemental Material~\cite{Supp:Mat}.
For the $\psi(4040)$, $Y(4360)$, and $\psi(4415)$, the two solutions are nearly identical and consistent within the present precision, and thus only one solution is reported.
The decay $\psi(3770)\to\SSm$ is observed for the first time with a significance of $5.5\sigma$, obtained by comparing the $\chi^2$ values of fits with ($\chi^2=20.6$) and without ($\chi^2=54.3$) the $\psi(3770)$ resonance, including systematic uncertainties via the covariance matrix.
No significant signals are found for $\psi(4040)$, $\psi(4160)$, $Y(4230)$, $Y(4360)$, $\psi(4415)$, or $Y(4660)$. Upper limits of the $\Gamma_{ee}{\cal{B}}$ for these charmonium(like) states decaying into the $\SSm$ final state are provided at the 90\% confidence level (CL) using a Bayesian approach \cite{Zhu:2008ca}. More details are summarized in the Supplemental Material~\cite{Supp:Mat}.

\begin{table}[htbp]
    \caption{Fit results for $\Gamma_{ee}\mathcal{B}$ ($10^{-3}$ eV) and $\phi$ (rad) with two solutions. Both statistical and systematic uncertainties are included. Here, the $\chi^2/{\rm n.d.f}$ indicates the fit quality, and ${\rm n.d.f.}$ denotes the number of degrees of freedom, defined as the number of data points minus the number of fixed parameters.}
    \centering
    \scalebox{0.92}{
        \begin{tabular}{l r@{ $\pm$ }l r c}
            \hline
            \hline
            Parameter                                      &\multicolumn{2}{c}{Solution I}         &\multicolumn{1}{c}{Solution II}       &$\chi^{2}/{\rm n.d.f}$     \\
            \hline
            $\phi_{\psi(3770)}$                              &$-2.0$   &$0.2$                       &$2.3  \pm 0.2~~$            &\multirow{2}{*}{$50.2/(52-4)$}  \\
            $\Gamma_{ee}\mathcal{B}_{\psi(3770)}$            &$75.3$   &$4.5$                       &$27.7 \pm 2.8~~$                                             \\

            $\phi_{\psi(4040)}$                              &$2.9$   &$0.7$                        &\multirow{2}{*}{$-~~~~~~~$} &\multirow{2}{*}{$46.2/(52-6)$}  \\
            $\Gamma_{ee}\mathcal{B}_{\psi(4040)}$            &$24.5$  &$14.6~(\textless 58.0)$                                                                  \\

            $\phi_{\psi(4160)}$                              &$-2.1$   &$0.1$                       &$-2.8 \pm 0.3~~$            &\multirow{2}{*}{$46.1/(52-6)$}  \\
            $\Gamma_{ee}\mathcal{B}_{\psi(4160)}$            &$56.9$   &$12.5~(\textless 81.0)$     &$9.5  \pm 7.5~~$                                             \\

            $\phi_{\psi(4230)}$                              &$-2.0$   &$0.1$                       &$-2.9 \pm 0.4~~$            &\multirow{2}{*}{$45.1/(52-6)$}  \\
            $\Gamma_{ee}\mathcal{B}_{\psi(4230)}$            &$42.0$  &$10.7~(\textless 61.8)$      &$5.3  \pm 3.9~~$                                             \\
   
            $\phi_{\psi(4360)}$                              &$-1.2$  &$0.4$                        &\multirow{2}{*}{$-~~~~~~~$} &\multirow{2}{*}{$38.3/(52-6)$}  \\
            $\Gamma_{ee}\mathcal{B}_{\psi(4360)}$            &$56.2$  &$97.2~(\textless 121.7)$                                                                   \\

            $\phi_{\psi(4415)}$                              &$-1.1$  &$0.2$                        &\multirow{2}{*}{$-~~~~~~~$} &\multirow{2}{*}{$39.7/(52-6)$}  \\
            $\Gamma_{ee}\mathcal{B}_{\psi(4415)}$            &$89.7$  &$20.3~(\textless 154.0)$                                                                   \\

            $\phi_{\psi(4660)}$                              &$-1.3$  &$0.2$                        &$-0.6 \pm 0.6~~$            &\multirow{2}{*}{$48.5/(52-6)$}  \\
            $\Gamma_{ee}\mathcal{B}_{\psi(4660)}$            &$51.4$  &$26.3~(\textless 98.1)$      &$7.6  \pm 11.3~~$                                             \\
            \hline
            \hline
        \end{tabular}}
        \label{tab:multisolution}
\end{table}

% \section{Summary} 
In summary, using  44 $\rm fb^{-1}$ of $\EE$ collision data collected with the BESIII detector at the BEPCII Collider, 
we present the first measurement of Born cross sections and EFFs for the $\EE\to\SSm$ reaction at 52 c.m.\ energies from 3.51 to 4.95~GeV. 
Fitting the dressed cross sections with a coherent sum of a continuum and individual charmonium(like) resonances, we observe $\psi(3770)\to\SSm$ for the first time with a significance of 5.5$\sigma$, including systematic uncertainties. 
The branching fraction of $\psi(3770)\to\Sigma^-\bar\Sigma^+$ is determined to have two solutions, $(2.9 \pm 0.3)\times 10^{-4}$ and $(1.1 \pm 0.1)\times 10^{-4}$, both of which are larger by at least an order of magnitude than the prediction based on scaling from the electronic width~\cite{Dobbs:2017hyd}. 
This tendency is consistent with that observed for its isospin partners, $\Sigma^+\bar\Sigma^-$~\cite{BESIII:2024umc} and $\Sigma^0\bar\Sigma^0$~\cite{BESIII:2024gql}, suggesting that the $B\bar{B}$ production might not originate purely from electromagnetic processes, but could involve contributions from charmonium(like) states.
Because the measured non-$D\bar{D}$ branching fraction of $\psi(3770)$ is unexpectedly large, about 15\%~\cite{BES:2008vad}, and remains inconsistent among different experiments~\cite{CLEO:2005mpm}, the first observation of a charmless $\psi(3770)$ decay into a hyperon pair sheds new light on the long-standing non-$D\bar{D}$ puzzle of $\psi(3770)$.
No significant signals are found for $\psi(4040)$, $\psi(4160)$, $Y(4230)$, $Y(4360)$, $\psi(4415)$, or $Y(4660)$, and 90\% CL upper limits on $\Gamma_{ee}{\cal{B}}$ are reported. These results provide crucial input for understanding the mechanisms governing vector charmonium(like) decays to $B\bar{B}$ final states and offer a new probe of exotic structures above the open-charm threshold. In addition, the measured ratios of Born cross sections among the three isospin-related channels $\Sigma^-\bar\Sigma^+$, $\Sigma^+\bar\Sigma^-$, and $\Sigma^0\bar\Sigma^0$ provide further validation of the VMD model
~\cite{Iachello:1972nu,Iachello:2004aq,Bijker:2004yu, Yang:2019mzq,Li:2021lvs}. Except for the region around $\sqrt{s} = 3.773$ GeV, where strong-interaction contributions may be present, the results at the other energy points are consistent with the VMD predictions within one to two standard deviations.

$Acknowledgments$. The BESIII Collaboration thanks the staff of BEPCII (https://cstr.cn/31109.02.BEPC) and the IHEP computing center for their strong support. This work is supported in part by National Key R\&D Program of China under Contracts Nos. 2025YFA1613900, 2023YFA1606000, 2023YFA1606704; National Natural Science Foundation of China (NSFC) under Contracts Nos. 12075107, 11635010, 11935015, 11935016, 11935018, 12025502, 12035009, 12035013, 12061131003, 12192260, 12192261, 12192262, 12192263, 12192264, 12192265, 12221005, 12225509, 12235017, 12342502, 12361141819;
the Fundamental Research Funds for the Central Universities No. lzujbky-2025-ytA05,  No. lzujbky-2025-it06,  No. lzujbky-2024-jdzx06;
the Natural Science Foundation of Gansu Province No. 22JR5RA389, No.25JRRA799;
the ``111 Center'' under Grant No. B20063;
the Chinese Academy of Sciences (CAS) Large-Scale Scientific Facility Program; the Strategic Priority Research Program of Chinese Academy of Sciences under Contract No. XDA0480600; CAS under Contract No. YSBR-101; 100 Talents Program of CAS; The Institute of Nuclear and Particle Physics (INPAC) and Shanghai Key Laboratory for Particle Physics and Cosmology; ERC under Contract No. 758462; German Research Foundation DFG under Contract No. FOR5327; Istituto Nazionale di Fisica Nucleare, Italy; Knut and Alice Wallenberg Foundation under Contracts Nos. 2021.0174, 2021.0299, 2023.0315; Ministry of Development of Turkey under Contract No. DPT2006K-120470; National Research Foundation of Korea under Contract No. NRF-2022R1A2C1092335; National Science and Technology fund of Mongolia; Polish National Science Center under Contract No. 2024/53/B/ST2/00975; STFC (United Kingdom); Swedish Research Council under Contract No. 2019.04595; U. S. Department of Energy under Contract No. DE-FG02-05ER41374.
~\\

$Data~availability$. There are no publicly available research data or software supporting this manuscript. Requests for further information or data should be sent to the authors.
%%%%%%%%%%%%%%%%%%%%%%%%%%%%%%%%%%%%%%%%%%%%%%%%%%%%%%%%%%%%%%%%%
%\bibliographystyle{apsrev4-1}
\bibliography{SigmamBCS_ref.bib}

\begin{widetext}
\begin{center}
\small
%% Saved at => 2025-09-11
M.~Ablikim$^{1}$\BESIIIorcid{0000-0002-3935-619X},
M.~N.~Achasov$^{4,c}$\BESIIIorcid{0000-0002-9400-8622},
P.~Adlarson$^{81}$\BESIIIorcid{0000-0001-6280-3851},
X.~C.~Ai$^{86}$\BESIIIorcid{0000-0003-3856-2415},
C.~S.~Akondi$^{31A,31B}$\BESIIIorcid{0000-0001-6303-5217},
R.~Aliberti$^{39}$\BESIIIorcid{0000-0003-3500-4012},
A.~Amoroso$^{80A,80C}$\BESIIIorcid{0000-0002-3095-8610},
Q.~An$^{77,64,\dagger}$,
Y.~H.~An$^{86}$\BESIIIorcid{0009-0008-3419-0849},
Y.~Bai$^{62}$\BESIIIorcid{0000-0001-6593-5665}
% ...
\\
\vspace{0.2cm}
(BESIII Collaboration)\\
\vspace{0.2cm} {\it
$^{1}$ Institute of High Energy Physics, Beijing 100049, People's Republic of China\\
% ...
$^{4}$ Budker Institute of Nuclear Physics SB RAS (BINP), Novosibirsk 630090, Russia\\
% ...
$^{31}$ INFN Sezione di Ferrara, (A)INFN Sezione di Ferrara, I-44122, Ferrara, Italy; (B)University of Ferrara, I-44122, Ferrara, Italy\\
% ...
$^{39}$ Johannes Gutenberg University of Mainz, Johann-Joachim-Becher-Weg 45, D-55099 Mainz, Germany\\
% ...
$^{62}$ Southeast University, Nanjing 211100, People's Republic of China\\
% ...
$^{64}$ State Key Laboratory of Particle Detection and Electronics, Beijing 100049, Hefei 230026, People's Republic of China\\
% ...
$^{77}$ University of Science and Technology of China, Hefei 230026, People's Republic of China\\
% ...
$^{80}$ University of Turin and INFN, (A)University of Turin, I-10125, Turin, Italy; (B)University of Eastern Piedmont, I-15121, Alessandria, Italy; (C)INFN, I-10125, Turin, Italy\\
$^{81}$ Uppsala University, Box 516, SE-75120 Uppsala, Sweden\\
% ...
$^{86}$ Zhengzhou University, Zhengzhou 450001, People's Republic of China\\

\vspace{0.2cm}
$^{\dagger}$ Deceased\\
% ...
$^{c}$ Also at the Novosibirsk State University, Novosibirsk, 630090, Russia\\
% ...
}
\end{center}
\end{widetext}

%%%%%%%%%%%%%%%%%%%%%%%%%%%%%%%%%%%%%%%%%%%%%%%%%%%%%%%%%%%%%%%%%
\end{document}